\documentclass[prd,twocolumn,aps,showpacs,preprintnumbers,amsmath,amssymb]{revtex4}
\usepackage{graphicx,graphics,color,epsfig}
\usepackage{dcolumn}
\usepackage{bm}
\usepackage{psfrag}

\begin{document}
\draft
\newcommand{\be}{\begin{equation}}
\newcommand{\ee}{\end{equation}}
\newcommand{\bea}{\begin{eqnarray}}
\newcommand{\eea}{\end{eqnarray}}
\def\lsim{\raise0.3ex\hbox{$\;<$\kern-0.75em\raise-1.1ex\hbox{$\sim\;$}}}
\def\gsim{\raise0.3ex\hbox{$\;>$\kern-0.75em\raise-1.1ex\hbox{$\sim\;$}}}
\def\Frac#1#2{\frac{\displaystyle{#1}}{\displaystyle{#2}}}
\def\no{\nonumber\\}
\def\slash#1{\ooalign{\hfil/\hfil\crcr$#1$}}
\def\ppbar{p\overline{p}}
\def\no{\nonumber}

\def\comm#1{\textcolor{black}{#1}}

%
\preprint{LAL-12-07}
\title{ Investigating charmonium production at LHC with the $\ppbar$ final state}
\author{
Sergey Barsuk, Jibo He, Emi Kou  and Benoit Viaud} 
\address{Laboratoire de l'Acc\'el\'erateur Lin\'eaire (UMR 8607), 
Universit\'e Paris-Sud XI, 91898 Orsay C\'edex, France}
\date{\today}
\begin{abstract}
We propose to investigate various charmonium states using their common decay channel to $\ppbar$ at LHC. 
Having the branching ratios for charmonium decaying into the $\ppbar$ final state measured or calculated, we propose to measure the charmonium production rate for both hadroproduction including soft-diffraction and inclusive production from $b$-hadron decays. We discuss the theoretical impacts in QCD of measuring different charmonium production rates and also the experimental prospects at LHCb, in particular, those for yet unmeasured $\eta_c$ and $h_c$. 
\end{abstract}
\maketitle

\section{Introduction}
The quarkonium, the bound state of heavy quarks, has been {playing an important role in understanding   the nature of the strong interaction}. 
The recent discoveries of the new type of heavy quark bound states, the so-called $XYZ$ particles, are further enriching the quarkonium physics (see~\cite{Brambilla:2010cs} for a recent review). 
The theoretical predictions for charmonium decay and production have been a great challenge of  the Quantum Chromo-Dynamics (QCD).  An important progress has been made by the effective field theory approach, called the Non-Relativistic QCD (NRQCD)~\cite{Bodwin:1994jh}. The NRQCD approach allows to systematically improve theoretical predictions by computing higher order terms in the expansion of $\alpha_s$ as well as of the velocity $v$.  While substantial theoretical efforts have been made, comparison of the NRQCD predictions to the experimental data still leaves open questions (see e.g.~\cite{Brambilla:2010cs, Ma:2010yw, Butenschoen:2012px,Artoisenet:2008fc}). 
In this article, we propose  to further investigate charmonium production as well as decay mechanisms at Large Hadron Collider (LHC) at CERN through a simultaneous measurement of various charmonium states using their decays into the $\ppbar$ final state. 

The LHC produces huge samples of primary and secondary charmonium, thanks to its extremely high luminosity as well as  the large $c\bar{c}$ and $b\bar{b}$ production cross sections of approximately 6 mb and 0.3 mb at $\sqrt{s}=7$ GeV energy, respectively~\cite{Aaij:2011jh}. 
 {The LHCb experiment is the main actor at heavy flavour studies at the LHC due to its precise vertex and track reconstruction, powerful particle identification and flexible trigger~\cite{Alves:2008zz}}. 
Within the LHCb acceptance,  
roughly $10^{12}$ $c\overline{c}$ pairs are produced and a  $10^8$ $J/\psi$  reconstructed through its $\mu^+\mu^-$ decay per 1 fb$^{-1}$ of data.

 The transverse momentum dependence of the production cross section as well as the angular distribution of the $J/\psi\to \mu^+\mu^-$ decay are known to carry important information to probe the NRQCD picture. 
Some discrepancies reported by the previous measurements of these observables at Tevatron~\cite{Abe:1997jz, Acosta:2004yw, Aaltonen:2009dm} motivate a further investigation at LHC. LHC brings a new information by measuring the production at a higher energy. 
In particular, the LHCb unique acceptance of $1.9<\eta <4.9$ fully instrumented coverage gives access to the QCD studies at the forward region~\cite{Aaij:2011jh}. Furthermore, powerful particle identification of LHCb makes it possible to  measure the production rates of different charmonium states. 

So far, charmonium study at LHC has been limited to using the $\mu^+\mu^-$ final state, thus mainly focused on $J/\psi$, $\psi(2S)$ or radiative transitions of $\chi_{c1,2}$ to $J/\psi$. 
\comm{However, reconstruction efficiency of the soft photon from $\chi_{c}$ decay leads to a dominant contribution to the systematic error of the measurement (see e.g.~\cite{2012tq}). The di-hadron final states at the hadron machines implies a huge combinatory background. The majority of the particles produced are $\pi$ and $K$. Thus, given charged hadron particle identification is available, this is the $\ppbar$ final state that gives a manageable background level. }
The $\ppbar$ final state is more convenient since most of the charmonium states can decay into it.  We propose to investigate charmonium prompt production as well as  inclusive charmonium yield from $b$-hadron decays using the $\ppbar$ final state, which may allow us to study all the charmonium states below the $D\overline{D}$ threshold, $J/\psi, \eta_c, \chi_{cJ}, h_c, \psi(2S), \eta(2S)$. We also suggest to test the potential $X(3872)\to \ppbar$ decay.  Prompt production assumes  hadroproduction including soft-diffraction, although we do not consider the latter in this article (see~\cite{HarlandLang:2009qe} for this subject).

While experimental measurements can provide only the information of the product of the cross section and the branching ratio, the branching ratios for the $\ppbar$ final state for many charmonia are rather well measured.  Using these measured values, we can extract the information of the production cross section. For $h_c$, the branching ratio for the $\ppbar$ final state is not known, and we estimate it theoretically below. 

In the next section, we propose the simultaneous measurement of various charmonium states with the $\ppbar$ final state at LHCb. We also discuss what new information in terms of testing the NRQCD can be obtained through this study. In section 3, we attempt to evaluate the branching ratio of $h_c\to\ppbar$ process. In section 4, we discuss other final states which could be used, and our conclusion is given in section 5. 

\section{
Investigating various charmonium states using the common decay mode into $\ppbar$}
If a given charmonium state has a significant branching ratio  
for the $\ppbar$ decay channel,  a simultaneous reconstruction of this charmonium state and the well-measured $J/\psi$ state via decay to $\ppbar$ is experimentally advantageous. 
\comm{For the topologically and kinematically similar channels, the systematic error, coming from efficiency calculations, detector description etc, cancels in the ratio. } 
In particular, relative prompt production and inclusive yield of charmonium state from $b$-hadron decays can be written as follows: 
\be
{\mathcal{R}}^{\rm prompt}_{\eta_c/\chi_{cJ}/h_c}\equiv 
\frac{\sigma (\eta_c/\chi_{cJ}/h_c) \times BR( (\eta_c/\chi_{cJ}/h_c) \to p \overline{p} )}
{\sigma ( J/\psi  ) \times BR( J/\psi  \to p \overline{p} )}
\ee
and 
\bea
&&\hspace{-0.5cm}{\mathcal{R}}^{b{\rm-inclusive}}_{\eta_c/\chi_{cJ}/h_c}\equiv  \\
&&
 \frac{BR (b\to (\eta_c/ \chi_{cJ}/ h_c)X) \times 
 BR( (\eta_c/ \chi_{cJ}/ h_c) \to p \overline{p} )}
{BR(b\to  J/\psi X ) \times BR( J/\psi  \to p \overline{p} )}. \no 
\eea
The $b$-inclusive decay assumes decays of $B$ mesons and $b$-baryons present according to their production fractions~\cite{Nakamura:2010zzi}. 
If the branching ratios $Br(( \eta_c/\chi_{cJ}/h_c) \to p \overline{p}) $ are known from other experiments or from theoretical computations, one can determine the production rate of different charmonium states produced promptly or from $b$-hadron decays. Indeed, such branching ratios are known except for $h_c$. In the next section, we estimate the missing branching ratio for $h_c\to \ppbar$.

On the experimental side, selection of the prompt charmonium decaying into $p \overline{p}$ pair candidates relies on searching for the high $p_T$ (anti-)protons, good quality $p \overline{p}$ vertex, and proton particle identification, with no  topologically clean handle to suppress the background. 
For prompt charmonium production, since minimum bias cross section is very large, around 60 mb~\cite{Aaij:2011er}, in order to retain significant charmonium samples via $\ppbar$ final states, the implementation of the dedicated trigger is important. 
For secondary charmonium coming from $b$ decays, additional efficient background suppression is achieved by requiring a significant $b$-flight distance. 

Projection from $J/\psi$ cross section measurement via $J/\psi\to \mu^+\mu^-$~\cite{Aaij:2011jh} and the trigger efficiency estimate suggest $10^{3}$ to $10^{4}$ of prompt $J/\psi\to\ppbar$ decays per 1 fb$^{-1}$ to be retained after trigger, reconstruction and selection at LHCb. {Assuming similar $\eta_c$ production cross section and background conditions, one can expect to observe $\eta_c$ both prompt and from $b$ decays, while a precise determination of the production rate $\sigma(\eta_c)$ or $Br(b\to\eta_c X)$ would still require a better information on  $Br(\eta_c\to \ppbar)$}. 

Measuring the cross section for different charmonium states has significant impacts on understanding the production mechanism of charmonium in  QCD. 
The well-established effective field theory framework of NRQCD  separates the short-distance part, which is set by the heavy-quark mass and is calculable by perturbative QCD, from the longer-distance part which is described by the universal matrix elements. 
While tremendous efforts have been made to improve the theoretical predictions confronted to the experimental measurements, it seems that a full explanation of the experimental data is still lacking \comm{(the size of the color-octet contribution is discussed e.g. in~\cite{Ma:2010yw, Butenschoen:2012px,Lansberg:2010cn})}. More observables, such as cross sections of different charmonium states proposed here, are certainly  welcome to seek for the missing pieces of the theoretical picture. More specifically, 
\begin{itemize}
\item 
The non-perturbative matrix elements used in NRQCD describing spin-singlet and spin-triplet states 
are related under heavy-quark spin symmetry. Therefore,  at least at the leading order, the matrix elements for $\eta_c$ and $h_c$ are related to those for $J/\psi$ and $\chi_c$, respectively. 
The measurement of the yet unknown  production  for $\eta_c$ and $h_c$: 
\be
\sigma(\eta_c), \sigma(h_c), Br(b\to X\eta_c), Br(b\to Xh_c)
\ee
 would provide a crucial test  of NRQCD. The theoretical predictions for these production rates can be found e.g. in~\cite{Biswal:2010xk,Sridhar:2008sc,Qiao:2009zg,  Bodwin:1992qr,Ko:1995iv,Beneke:1998ks}.  
 \item
The theoretical predictions for charmonium production rate include only the {\it direct prompt} production. On the other hand, experimentally, promptly produced charmonium states are identified as coming from the primary vertex. Thus, by definition, they also include a  feed-down from  higher charmonium states. For example, experimentally measured prompt $J/\psi$ production rate comprises also the feed-down contributions from $\psi(2S)\to J/\psi \pi\pi $ and $\chi_{cJ}\to J/\psi \gamma$. 
Therefore, a more complete information of the production rates of the whole charmonium system is necessary to yield an unambiguous theory to experiment comparison.
\item 
Measuring the ratio of different charmonium production is advantageous. 
Experimentally, the systematic effects partially cancel in the ratio. 
Theoretically, such a ratio is often easier to predict than the actual magnitude of the cross sections. Moreover, it has been pointed out~\cite{Brambilla:2010cs} that in some cases, theoretical uncertainties  coming e.g.  from the renormalisation and the factorisation scales may also cancel. 
\item The long-distance matrix element in NRQCD is supposed to be universal for any production and decay processes. A verification of this statement using various charmonium decays and productions is essential.  
\end{itemize}

\section{Charmonium decays into the $\ppbar$ final state}
\begin{table*}
\begin{center}
\begin{tabular}{|c||c|c|c|c|c|c|c|}
\hline\
&$\ppbar$&$\Lambda\overline{\Lambda}$ &$\overline{\Xi}^+{\Xi}^-$ &$ \phi K^+K^-$ &$ \phi \pi^+\pi^-$ & $\eta_c\gamma $& $\phi\phi$ \\
\hline \hline 
$J/\psi$ &$2.17\pm0.07$&$1.61\pm 0.15 $&$0.85\pm 0.16$&$1.83\pm 0.24$&$0.87\pm 0.08$&$17\pm 4$& forbidden\\
$\eta_c$ &$1.3\pm0.4$&$1.04\pm 0.31$&unknown&$2.9\pm 1.4$&unknown&forbidden &$2.7\pm 0.9$\\
$\chi_{c0}$ &$	0.223 \pm 0.013 $&$0.33\pm 0.04$&$0.49\pm 0.07$&$0.98 \pm 0.25 $&unknown& forbidden&$0.91\pm 0.19$\\
$\chi_{c1}$ &$0.073\pm0.004$&$ 0.118\pm 0.019$&$0.084 \pm 0.023 $&$0.43\pm 0.16$&unknown& forbidden&unknown\\
$\chi_{c2}$ &$0.072\pm0.004$&$0.186\pm 0.027$&$0.155 \pm 0.035 $&unknown&unknown& forbidden&$1.48\pm 0.28$\\
$\psi(2S)$ &$0.276\pm0.012$&$0.28\pm 0.05 $&$0.18\pm 0.06$&$0.070\pm 0.016$&$ 0.117\pm 0.029$&$3.4\pm 0.5$& forbidden\\\hline

\end{tabular}
\end{center}
\caption{Measured branching ratio ($\times 10^{3}$) for the charmonium decaying to the hadronic final state~\cite{Nakamura:2010zzi}, which allows simultaneous measurement of the production rate for  several charmonium states.  }
\label{table}
\end{table*}%

Among a hundred of possible decay channels of charmonium, the charge ($C$) and G-parity ($G$) conservations make only few  decay channels allowing a simultaneous measurement of charmonia with different  quantum numbers $J^{PC}=\{1^{--}, 0^{-+}, 0^{++}, 1^{++}, 2^{++}, 1^{+-}\}$. 
For example, final states with two pseudoscalars, two vectors, one vector plus one pseudoscalar are not suitable for the simultaneous measurement of all charmonium states. 
From experimental point of view, it is important to keep in mind requirements specific to hadron machines: avoid as much as possible final states with neutral particles $\pi^0/\gamma$ and  $K^0$ which have low reconstruction efficiency and/or introduce large combinatorial background. 
These considerations suggest the $\ppbar$ final state, which is a rather simple two body process, to be ideal. In the next section, we discuss other final states potentially promising for simultaneous measurement of some charmonium states.

The  branching ratios for the decays to the $\ppbar$ final state have been measured for many charmonia as shown in Table~\ref{table}~\cite{Nakamura:2010zzi}. 
In the following, we attempt to estimate the branching ratio for $h_c\to\ppbar$ which has not been observed yet. 
The $h_c$ had been a "missing charnomium" for a long time. While by now, the $e^+e^-$ machines such as CLEO and BESIII have accumulated a significant sample of $h_c$~\cite{Ablikim:2010rc,Dobbs:2008ec}, only two decay modes have been observed and a little information is known about this state. 
Thanks to the prolific $c\overline{c}$ production, LHC is delivering huge number of the $h_c$, making it possible to access other $h_c$ decays. 
It should also be noted that the exclusive $B\to Kh_c$ is the so-called factorisation forbidden process and an observation of this channel is also very important to have a better control of various corrections in the theoretical computations of hadronic $B$ decays~\cite{Beneke:2008pi}. 

In early 80's, a proposal based on the perturbative QCD  was made for computing the charmonium decaying into baryon and anti-baryon, where interesting power counting rule as well as the so-called helicity selection rule are derived~\cite{Brodsky:1981kj}. This selection rule forbids such processes as $\eta_c/\chi_{c0}/h_c\to\ppbar$ at the collinear and massless limit. On the other hand,   the observed branching ratios for the first two decay channels indicate an importance of non-leading contributions. Various theoretical efforts have been made to include these contributions while the whole picture is not fully clarified yet. Thus, in the following,  we attempt to obtain an estimate of the $h_c\to \ppbar$ branching ratio based on simple arguments instead of using a more elaborated QCD picture. 

We write the amplitude for $h_c\to$ hadrons (hadrons to be $\ppbar$ here) by ''factorising'' the initial and the final state as: 
\begin{equation}
{\mathcal{A}}(h_c\to {\rm hadrons}) = 
{\mathcal{A}}(h_c\to ggg)\times {\mathcal{A}}((ggg)\to{\rm hadrons}) 
\end{equation}
The first part of the right hand side can be related to the total hadronic width of $h_c$. 
 \comm{It has been theoretically estimated~\cite{Suzuki:2002sq,Godfrey:2002rp} that the $h_c$ decay width is shared approximately equally by the radiative  $h_c\to \eta_c\gamma$ decay  and hadronic decays, i.e. $\Gamma(h_c\to\ {\rm hadrons})=530\pm 80 {\rm \ KeV}$, $\Gamma(h_c\to \eta_c \gamma)=520\pm 90 {\rm \ KeV}$ ~\cite{Suzuki:2002sq}.}
  Indeed, the former was recently measured as: $Br(h_c\to \eta_c\gamma)=( 53 \pm 7 ) \%$~\cite{Ablikim:2010rc,Dobbs:2008ec}. 
In order to estimate the part $(ggg)\to {\rm hadrons}$, we utilize the well-measured $J/\psi$ hadronic decay. The $J/\psi$ hadronic decay within the same approximation can be written as: 
\begin{equation}
{\mathcal{A}}(J/\psi\to{\rm hadrons}) = 
{\mathcal{A}}(J/\psi\to ggg) \times {\mathcal{A}}((ggg)\to{\rm hadrons})  \label{eq:jpsifac}
\end{equation}
The left hand side for hadrons=$\ppbar$ and the hadronic width  can be extracted from the experimental data for $J/\psi$. Then, taking into account the different quantum numbers between 3 gluons in $1^{+-}$ state and in ${1^{--}}$, one can readily  obtain the relation between these amplitudes, 
which leads to: 
\be
Br(h_c  \to  p\bar{p}) \simeq (3.2\pm 0.5)\times 10^{-3}. 
\ee 
The quoted error reflects only the uncertainty of the experimental input~\cite{Nakamura:2010zzi}: branching ratios for $J/\psi\to ggg$, $J/\psi\to \ppbar$ and $h_c \to ggg$, where the last one assumes $Br(h_c\to ggg)= 1-Br(h_c\to \eta_c\gamma)$. 
Error introduced by the assumption mentioned above is not taken into account. 
In naive QCD estimate, one expects the ratio $Br(J/\psi\to\ppbar)/Br(h_c\to\ppbar)$ to respect the power counting law, scaling as $(M_{h_c}/M_{J/\psi})^n$. For the simple example, $Br(J/\psi\to\ppbar)/Br(\psi(2S)\to\ppbar)$, $n\simeq 8$ is obtained. On the other hand, with the helicity forbidden channel $h_c\to\ppbar$, various corrections have to be taken into account. 
\comm{The obtained value is an order of magnitude larger than an estimate from QCD~\cite{Murgia:1996bh} 
while is consistent with the other estimates~\cite{Kuang:1988bz,Liu:2010um} (note that the decay rate given in~\cite{Kuang:1988bz} is smaller but if the same hadronic decay width is applied, the two results become consistent).}

\section{Other common final states for charmonium decays}

\subsection{Other baryonic channels}
Given significant branching ratios for several charmonia decaying into other baryon final states, 
they can also be used for measuring production rates. The following channels have been suggested: 
\bea
{\rm charmonium}\ &\to&
\Lambda\overline{\Lambda}\to (p\pi^-)(\overline{p}\pi^+) \no \\
{\rm charmonium}\ &\to&\overline{\Xi}^+{\Xi}^-\to (\overline{\Lambda}\pi^+)({\Lambda}\pi^-)\no \\
&\ & \to ((\overline{p}\pi^+)\pi^+)((p\pi^-)\pi^-). \no
\eea
The measured branching ratios for charmonium $\to\Lambda\overline{\Lambda}$ and $\overline{\Xi}^+\Xi^-$  are similar to the ones for the $\ppbar$ final state (see Table~\ref{table}). 
{Experimentally, reconstruction efficiency reduces because 4 (6) tracks in the final state have to be reconstructed; trigger efficiencies reduce because of lower particle momenta. On the other hand, 2 or 4 secondary vertices in the event provide a clean signature that allows to strongly suppress combinatorial background.} 
Thus, these processes could be complementary to the $\ppbar$ final state. 
The $h_c$ branching ratios in the same method as in the previous section are estimated  to be of the order $10^{-3}$ similar to the $\ppbar$ final state. 
 
\subsection{The $\phi K^+K^-,\ \phi \pi^+\pi^-$ channels}
The $\phi K^+K^-$ and $\phi \pi^+\pi^-$ channels could be also produced from all the charmonium spin states, since the $K^+K^-/\pi^+\pi^-$ part may come from resonance states with different spin and/or continuum. Using the decay of $\phi\to K^+K^-$, this channel could be experimentally accessible at LHC. In addition, if $K^+K^-/\pi^+\pi^-$ pair comes from narrow $f_X$ resonance with defined quantum numbers, the charmonium decay reconstruction becomes even more feasible. 

\subsection{The $\eta_c\gamma$ channel for $C$ odd charmonium}
An example that can provide simultaneous measurement for $C$ odd charmonia is the $\eta_c\gamma$ final state. It allows the ratio measurement of e.g. $J/\psi$ and $h_c$. 

In the case of using the $\ppbar$ invariant mass spectrum, the experimental resolution should allow resolving the $\chi_{c1}$ and $h_c$ states having a 15 MeV mass difference. The advantage of the $\eta_c\gamma$ final state is the absence of the decays from $\chi_{cJ}$,  so that the $h_c$ contribution can be unambiguously identified. 

A priori, this channel is not ideal for LHC because  the reconstruction efficiency of a low energy photon is small.  Nevertheless,  when charmonium is produced from the secondary vertex, i.e. from $b$-hadrons,
the requirement of the secondary vertex can give necessary background suppression. 
We are particularly interested in this channel since this final state carries a half of the $h_c$ branching ratio as mentioned above. The estimates suggest, that  
$h_c$ may be reconstructed with one LHCb nominal year luminosity of $2 fb^{-1}$  by using the subsequent  $\eta_c\to \phi\phi $ channel whose branching ratio is known as $Br(\eta_c\to \phi\phi )=(2.7\pm 0.9)\times 10^{-3} $~\cite{Nakamura:2010zzi}.

\subsection{The $\phi\phi$ channel for $C$ even charmonium }
The $C$ even charmonium, such as  $\eta_c$ and $\chi_{c0,1,2}$, decaying to $\phi \phi$ is particularly suitable for LHCb 
thanks to the large branching ratio of $\phi \to K^+K^-$ and clean signature provided by two narrow $\phi$ signals. 
A measurement of the S-wave spin-singlet charmonium $\eta_c$  has a very important consequence in having deeper insight into the NRQCD picture, while it has not been observed due to a lack of an easy decay channel to look for at the hadron colliders. 
Together with $\ppbar$ final state, the $\phi\phi$ channel should be useful for further investigating $\eta_c$ production. 

\section{Conclusions}
We proposed to investigate various charmonium states using their common decay channel to the $\ppbar$ final state at LHC. So far, the charmonium studies at hadron machines have been limited to $J/\psi$ and  $\psi(2S)$ 
via their decay into the $\mu^+\mu^- $, which is the cleanest way to reconstruct them at hadron machines, and $\chi_{cJ}$ reconstructed through $J/\psi \gamma$. 
Powerful particle identification of LHCb makes it possible to use the $\ppbar$ final state to which also other charmonium states can decay. 

Simultaneous reconstruction of various charmonium states and the well-measured $J/\psi$ state via their decay to $\ppbar$ significantly reduces the experimental systematic uncertainty thanks to measuring the ratio of topologically identical channels.  Such simultaneous investigation will allow us to access the production rates for the yet unmeasured charmonium states, such as $\eta_c$ and $h_c$. We discussed the theoretical impact of these new measurements, in particular, on improving our understanding of production mechanisms in QCD. It should be also emphasized a potential 
 observation of $h_c$ with the $\ppbar$ final state at LHC,  since $h_c$ has only recently been discovered and  little is known on this state so far. 
Our study will be also extended to higher mass states such as $\psi(2S), \eta_c(2S)$ and furthermore to yet un-interpreted $X(3872)$ state. An observation of $X(3872)$ into the $\ppbar$ final state would add an important clue in theoretical  identification of this particle (see~\cite{Braaten:2007sh} for recent estimates of $Br(X(3872)\to\ppbar)$ in the case of the molecule interpretation). 

We also investigated other final states, such as $\Lambda\overline{\Lambda}$, $\overline{\Xi}^+\Xi^-$, $\phi KK$, $\phi\pi\pi$, $\phi\phi$ and $\eta_c\gamma$, which can be complementary to the $\ppbar$ final state for systematic  charmonium production studies.  

In addition, we proposed to use hadronic final states, in particular $\ppbar$, for measuring the inclusive yield of charmonium states from $b$-hadron decays. 

In 2010, using 36 pb$^{-1}$ of data, LHCb has observed $J/\psi\to \ppbar$ decay, which is the first hadron final state of prompt produced charmonium decay reconstructed at hadron machines~\cite{WB}. 
In 2011, LHCb has recorded the integrated luminosity of more than 1 fb$^{-1}$ proving a stable operation at the designed luminosity of 2-4 $\times$ $10^{32}$ cm$^{-2}$s$^{-1}$. 
With the 2011 data, a significant result on $\eta_c\to \ppbar$ both for prompt production and for inclusive $b$-handron decays can be expected.  

\comm{In the longer time scale, with the increased LHC energy and thus, the $b\bar{b}$ cross section,  bottomonium decaying into $\ppbar$ can also be searched for. This will be an important QCD test while at present only $\Upsilon (1S)\to \ppbar$ is observed. }


\bigskip
\section*{Acknowledgments}
The authors would like to thank  Jacques Lefran\c cois for sharing his idea of using the charmonium decaying into baryonic $\Lambda\overline{\Lambda}$ and $\overline{\Xi}^+\Xi^-$ final states.  S.B. and E.K.  acknowledge useful comments by John Rosner concerning the $h_c$ state. The work by E.K. was supported in part by the ANR contract “LFV-CPV-LHC” ANR-NT09-508531.


\end{document}